\documentclass[runningheads]{llncs}
\usepackage[T1]{fontenc}
\usepackage{amsfonts,amsmath,amssymb}
\usepackage[font=footnotesize,labelfont=bf, skip=2pt]{caption}
\usepackage{rotating}
\usepackage{graphicx}
\usepackage{color}
\usepackage{hyperref}
\usepackage{graphicx}
\usepackage{comment}
\usepackage{csquotes}
\newcommand{\orcid}[1]{
	\href{https://orcid.org/#1}{\includegraphics[scale=0.4]{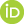}}
}
\usepackage{fontawesome}  
\newcommand{\attr}{n}

\newcommand{\event}{\ensuremath{e}}
\newcommand{\eventLog}{\ensuremath{L}}

\newcommand{\natNum}{\ensuremath{\mathbb{N}}}

\newcommand{\org}{\ensuremath{o}}
\newcommand{\orgs}{\ensuremath{\mathcal{O}}}

\newcommand{\univAttr}{\ensuremath{\mathcal{N}}}

\newcommand{\univEvents}{\ensuremath{\mathcal{E}}}

\newcommand{\universe}{\mathcal{U}}
\newcommand{\multiset}{\mathbb{B}}
\newcommand{\outputt}{t \bullet}
\newcommand{\inputt}{\bullet t}
\newcommand{\outputp}{p \bullet}
\newcommand{\inputp}{\bullet p}
\newcommand{\noMove}{\gg}
\newcommand{\prvLog}{L^{prv}}
\newcommand{\pubLog}{L^{pub}}

\begin{document}
\title{Federated Conformance Checking}
%
%\titlerunning{Abbreviated paper title}
% If the paper title is too long for the running head, you can set
% an abbreviated paper title here

\author{Majid Rafiei\orcid{0000-0001-7161-6927}\textsuperscript{\href{mailto:majid.rafiei@pads.rwth-aachen.de}{\faEnvelopeO}} \and
    Mahsa Pourbafrani\orcid{0000-0002-7883-1627} \and
	Wil M.P. van der Aalst\orcid{0000-0002-0955-6940}}
\authorrunning{M. Rafiei et al.}
% First names are abbreviated in the running head.
% If there are more than two authors, 'et al.' is used.
%
\institute{Chair of Process and Data Science, RWTH Aachen University, Aachen, Germany \\
% \email{\{majid.rafiei,wvdaalst\}@pads.rwth-aachen.de} 
}

\maketitle              % typeset the header of the contribution
\begin{abstract}
Conformance checking is a crucial aspect of process mining, where the main objective is to compare the actual execution of a process, as recorded in an event log, with a reference process model, e.g., in the form of a Petri net or a BPMN. Conformance checking enables identifying deviations, anomalies, or non-compliance instances. It offers different perspectives on problems in processes, bottlenecks, or process instances that are not compliant with the model. %It provides valuable insights into process inefficiencies, bottlenecks, or non-compliant behavior.
Performing conformance checking in federated (inter-organizational) settings allows organizations to gain insights into the overall process execution and to identify compliance issues across organizational boundaries, which facilitates process improvement efforts among collaborating entities.
In this paper, we propose \textit{a privacy-aware federated conformance-checking approach} that allows for evaluating the correctness of overall cross-organizational process models, identifying miscommunications, and quantifying their costs. 
For evaluation, we design and simulate a supply chain process with three organizations engaged in purchase-to-pay, order-to-cash, and shipment processes. We generate synthetic event logs for each organization as well as the complete process, and we apply our approach to identify and evaluate the cost of pre-injected miscommunications. 

\end{abstract}

\section{Introduction}
Process mining encompasses a range of techniques aimed at discovering, analyzing, and improving business processes \cite{van2016process}. By leveraging event logs, process mining provides evidence-based and actionable insights into the actual execution of processes. The field of process mining comprises three fundamental types of analysis: (1) \textit{process discovery}, where the goal is to learn real process models from event logs, (2) \textit{conformance checking}, where the aim is to find commonalities and disconformities
between a process model and the corresponding event log, and (3) \textit{process enhancement}, where the aim is to extend or improve process models using different aspects of the available data. In this paper, our main focus is on conformance checking, which plays a pivotal role in process mining. It enables the detection of deviations, anomalies, and instances of non-compliance. 

The sub-discipline of process mining focusing on the collaborative discovery, monitoring, analysis, and improvement of cross-organizational processes is referred to by various terms such as \textit{inter-organizational process mining}, \textit{cross-organizational process mining}, and \textit{federated process mining} \cite{WilIOPM,EDI_IOPM,Performance_IOPM,Wil_federated}.
In inter-organizational collaborations, organizations work together to achieve shared goals, often requiring the exchange of information, goods, or services across organizational boundaries. Conformance checking, in this context, offers organizations insights into the overall execution of processes and facilitates the identification of compliance issues that span organizational boundaries. This, in turn, aids in driving process improvement efforts among collaborating entities, leading to enhanced operational efficiency and alignment.
%However, performing conformance checking in federated or inter-organizational settings presents several challenges due to its distributed nature and the need to share and analyze sensitive data. It raises privacy concerns where the conformance checking should be done without revealing the whole internal process of each involved organization in the inter-organizational process.
However, due to its distributed nature and the need to share and analyze sensitive data, performing conformance checking in federated or inter-organizational settings presents several challenges. It raises privacy concerns given that conformance checking should be done without revealing the entire internal process of each involved organization in the overall inter-organizational process, while also taking into account the entire process and communication points. 

In this paper, we propose a privacy-aware federated conformance-checking approach. Our approach addresses privacy concerns and enables organizations to perform conformance checking while safeguarding sensitive information. %By evaluating the correctness of cross-organizational process models, identifying and quantifying miscommunication costs, and computing the overall fitness of cross-organizational process instances, our approach provides a comprehensive analysis of process conformance in federated environments.
Our approach provides a comprehensive analysis of process conformance in federated environments by evaluating the correctness of cross-organizational process models, identifying and quantifying miscommunication costs, and computing the overall fitness of cross-organizational process instances.

To validate the effectiveness of our proposed approach, we simulate a supply chain process involving three organizations. Specifically, we focus on purchase-to-pay, order-to-cash, and shipment processes, which are critical components of the supply chain domain. By designing this synthetic supply chain process, we have full control over its characteristics, allowing us to evaluate our approach in a controlled environment.
We generate synthetic event logs, capturing the activities and interactions within the supply chain process. These event logs serve as the basis for our conformance-checking analysis. By applying our proposed approach to these synthetic event logs, we can assess the accuracy of our method for detecting and quantifying pre-injected miscommunications.

The remainder of this paper is structured as follows. In Section~\ref{sec:related_work}, we discuss related work. In Section~\ref{sec:preliminaries}, the preliminaries are explained. Section~\ref{sec:approach} provides details of the proposed approach. In Section~\ref{sec:evaluation}, we evaluate our approach, and Section~\ref{sec:conclusion} concludes the paper.

\section{Related Work}\label{sec:related_work} 
In this section, we present an overview of the research conducted in the field of federated process mining, which is also known as inter-organizational process mining or cross-organizational process mining. We reference previous studies that have explored the privacy and confidentiality aspects related to this discipline.

\subsection{Federated Process Mining}
In \cite{WilIOPM}, inter-organizational process mining is explained, and various categories of inter-organizational data flows are characterized. In \cite{EDI_IOPM}, the authors utilize EDI messages to illustrate an effective case study of inter-organizational process mining in the automobile industry. In \cite{Performance_IOPM}, the authors focus on enhancing process performance by leveraging insights gained from cross-organizational process mining. In \cite{Joos_IOPM}, an approach is proposed to compare collections of process models and event logs recorded in different Dutch municipalities.
In \cite{artifact_driven}, an artifact-driven approach for process monitoring is introduced. This approach exploits the Internet of Things (IoT) paradigm to monitor business process by capturing and analyzing the status of physical artifacts. Although such an approach can go beyond the scope of a single organization to monitor business processes, it highly depends on sensor equipped physical artifacts and ignores the message passing phase of collaboration, which is essential to capture inter-organizational miscommunications.

The potential of cloud computing \cite{WilCloud} and blockchains \cite{blockchain_IOPM} has also been acknowledged within the context of inter-organizational process mining. In \cite{SCM_IOPM}, the authors present an approach for discovering distributed processes in supply chains. In \cite{Pertinet_IOPM}, the authors describe fundamental patterns to capture modeling concepts commonly encountered in supply chains. 
In \cite{cross_org_workflow}, a process mining approach is introduced to uncover coordination patterns and workflow models from resource allocation logs of different organizations. Then, a process integration approach combines these models and coordination patterns to create a cross-organizational workflow model.
In \cite{cchp}, the authors propose an algorithm to discover Intra-department Healthcare Process (IHP) models based on different patterns among medical departments.
In \cite{cdcbpm}, a formal method is proposed to systematically model and verify cross-department processes, taking into account various coordination patterns among different departments.
Finally, in \cite{Wil_federated}, a framework is introduced that recommends event log abstractions to enable cross-organizational process mining.

\subsection{Privacy-Aware Federated Process Mining}
Recent research in this field has increasingly focused on addressing privacy and confidentiality issues, which are widely recognized as the primary barriers to federated process mining. In \cite{smcProcessMining}, a technique based on secure multi-party computation is proposed for discovering directly follows graphs, which focuses on two parties.
In \cite{liu2019towards}, the authors propose a framework for sharing public process models and discovering organization-specific process models from multiple parties, relying on a trusted third party. In \cite{cross_silo}, the authors suggest an approach for discovering process models in inter-organizational settings, utilizing differential privacy and secure multi-party computation. In \cite{rafiei_federated_process}, the authors propose an abstraction-based approach for privacy-aware federated process mining that focuses on process discovery. 
In \cite{trusted_exec}, the authors present an approach that enables process discovery on multi-actor event data while preserving the confidentiality and integrity of the original records in an inter-organizational business context. This approach prioritizes data confidentiality at the organizational level and imposes computational overhead on mining algorithms.

While previous work primarily focused on process discovery in process mining, our work, to the best of our knowledge, is the first to propose an approach for federated conformance checking, taking into account privacy and confidentiality concerns.

\section{Preliminaries}\label{sec:preliminaries}
In this section, we provide definitions of event logs, Petri nets, and alignments that are used in the rest of the paper.
We start with introducing basic notations and mathematical concepts.
Let $A$ be a set. $B(A)$ is the set of all multisets over $A$. 
Given $A$ and $B$ as two multisets, $A \uplus B$ is the sum over multisets, e.g., $[a^2,b^3] \uplus [b^2,c^2] = [a^2,b^5,c^2]$,
and $A\setminus B$ is the multiset (set) difference, e.g., $[a^2,b^3] \setminus [b^2,c^2] = [a^2,b]$. 
We define a finite sequence over $A$ of length $n$ as $\sigma {=} \langle a_1, a_2,\dots, a_n\rangle$ where $\sigma(i) {=} a_i {\in} A$ for all $i {\in} \{1,2,\dots,n\}$. The set of all finite sequences over $A$ is denoted with $A^*$. 
% and the set of all elements of $\sigma$ is written as $\{a {\in} \sigma\}$.

\subsection{Event Log}
An event log is a collection of events. An event corresponds to an activity execution and can be characterized by various attributes, e.g., an event often has a timestamp attribute referring to the time at which the event happened. 

\begin{definition}[Event]
	\label{def:event}
    Let $\univEvents$ be the universe of event identifiers and $\univAttr$ be the universe of attribute names.
    For any $e \in \univEvents$ and attribute name $n \in \univAttr$, $\pi_n(e) \in \universe_n$ is the value of attribute $n$ for event $e$, where $\universe_n$ is the universe of attribute values for attribute $\attr$. If $e$ does not have an attribute named $n$, then $\pi_n(e)= \bot~(null)$. 
    We assume the following attributes to always have a value for any event $\event$:
    \vspace{-0.15cm}
    \begin{itemize}
        \item $\pi_{cid}(\event) \in \universe_{cid}$; capturing the \emph{case identifier} of the process instance,
        \item $\pi_{act}(\event) \in \universe_{act}$; capturing the  \emph{activity} executed, and
        \item $\pi_{time}(\event) \in \universe_{time}$; capturing the  \emph{timestamp} at which the event happened.
    \end{itemize}
\end{definition}

\begin{definition}[Event Log]
	\label{def:event_log}
    Let $\univEvents$ be the universe of events. An event log $\eventLog$ is a collection of events, i.e., $\eventLog \subseteq \univEvents$. 
\end{definition}

Table~\ref{tbl:sample_event_log} shows a fragment of an event log recorded by a manufacturer's information system by executing a purchase-to-pay process, where case identifiers refer to order numbers.

\begin{table}[t]
\caption{A fragment of an event log recorded by a manufacturer's information system by executing a purchase-to-pay process.}
\centering
\scriptsize
\begin{tabular}{l|l|l}
\hline
cid & act             & time   \\ \hline
c1            & create purchase order ($po$) & 01.01.2023-08:00:00  \\ 
c1            & send order request ($so$)    & 01.01.2023-08:30:00 \\ 
c1            & goods receipt ($gr$)         & 02.01.2023-09:00:00 \\ 
c1            & invoice receipt ($ir$)       & 02.01.2023-09:30:00 \\ 
c1            & payment ($pa$)               & 02.01.2023-11:00:00 \\ 
c2            & create purchase order ($po$) & 02.01.2023-11:30:00 \\ 
c2            & send order request ($so$)    & 02.01.2023-11:40:00 \\ 
c2            & invoice receipt ($ir$)       & 02.01.2023-16:30:00 \\ 
c2            & goods receipt ($gr$)         & 02.01.2023-17:30:00 \\ 
c2            & payment ($pa$)               & 03.01.2023-10:30:00 \\ 
c3            & create purchase order ($po$) & 03.01.2023-10:40:00 \\ 
c3            & send order request ($so$)    & 03.01.2023-10:50:00 \\ 
c3            & invoice receipt ($ir$)       & 03.01.2023-16:30:00 \\ 
c3            & payment ($pa$)         & 03.01.2023-17:00:00 \\ 
c3            & goods receipt ($gr$)               & 03.01.2023-17:30:00\\ \hline
\end{tabular}
\label{tbl:sample_event_log}
\end{table}

\begin{definition}[Trace, Simple Event Log]
	\label{def:trace}
    \sloppy{Let $\eventLog \subseteq \univEvents$ be an event log. $cases(L)=\{\pi_{cid}(e)|e\in L\}$ is the set of case identifiers, and $events(L,c)=\{e\in L|\pi_{cid}(e)=c\}$ is the set of events in case $c$.} $trace(L,c)=\langle e_1,...,e_n\rangle \in \eventLog^*$ is a trace such that $\{e_1,...,e_n\}=events(L,c)$, and $\forall_{1\leq i<n}\ e_i{\prec}\ e_{i+1}$, where ${\prec}$ is a total order on events such that $\forall_{e_1, e_{2}\in L}\ e_1{\prec}e_2 \implies \pi_{time}(e_1)\leq \pi_{time}(e_2)$.
    $variant(L,c) = \langle \pi_{act}(e_1),\pi_{act}(e_2), ..., \pi_{act}(e_n)\rangle \in \universe_{act}^*$ is called the corresponding trace variant of $trace(L,c)$.
    $\bar{L}=\{trace(L,c)|c\in cases(L)\}$ is the set of traces in $L$ and $\tilde{L}=[variant(L,c)|c\in cases(L)]$ is the multiset of trace variants in $L$, which is called a \textit{simple event log}.
\end{definition}

\begin{definition}[Applying Functions to Sequences]\label{def:funseq}
    Let $f: X \nrightarrow Y$ be a partial function.
    $f$ can be applied to sequences of $X$ using the following recursive definition:
    (1) $f(\langle~\rangle) = \langle~\rangle$ and (2) for $\sigma \in X^*$ and $x\in X$:
    $$f(\langle x \rangle \cdot \sigma) = \begin{cases}
    f(\sigma) & \mbox{if} \ x \not\in dom(f)\\
    \langle f(x) \rangle \cdot f(\sigma) & \mbox{if} \ x \in dom(f)
    \end{cases}$$
\end{definition}

\subsection{Petri Nets}

A Petri net is represented by a directed graph, with nodes representing \textit{places} and \textit{transitions}, and edges representing the flow of \textit{tokens}\footnote{\scriptsize Tokens are represented as black dots.} between places and transitions.
Each transition is represented by a square, and each place is represented by a circle. Transitions are used to model activities, and they are connected via places that model possible states of the process. The state of a Petri net is determined by the distribution of tokens over places and is referred to as its \textit{marking}.

\begin{definition}[Petri Net]
	\label{def:petri_net}
	A Petri net is a triplet $N = (P,T,F)$, $P$ is a finite set of places, $T$ is a finite set of transitions which it allows $P \cap T = \emptyset$, and $F \subseteq (P \times T) \cup (T \times P)$ is a set of directed arcs. A marked Petri net is a pair $(N,M)$, where $N=(P,T,F)$ is a Petri net and $M \in \multiset(P)$ is a multiset over places denoting the markings of the net. 
\end{definition}

Given a Petri net $N = (P,T,F)$, for any $x \in P \cup T$, $\bullet x = \{ y \mid (y,x) \in F \}$ denotes the set of input nodes and $x \bullet = \{ y \mid (x,y) \in F \}$ denotes the set of output nodes. A transition $t \in T$ is enabled in marking $M$ of net $N$, denoted as $(N,M)[t\rangle$, if each of its input places $\inputt$ contains at least one token.
%An enabled transition $t$ may fire, i.e., thereby consuming one token from each input place in $\inputt$ and producing one token for each output place in $\outputt$. 
It is possible for an enabled transition $t$ to occur, in which case one token will be consumed from each input location in $\inputt$ and one token will be produced for each output place in $\outputt$. To represent with formal notations, the marking resulting from firing enabled transitions $t$ in marking $M$ of Petri net $N$ is $M' = (M \setminus \inputt) \uplus \outputt$.   %Formally: $M' = (M \setminus \inputt) \uplus \outputt$ is the marking resulting from firing enabled transition $t$ in marking $M$ of Petri net $N$.
$(N,M)[t\rangle(N,M')$ denotes that $t$ is enabled in $M$ and firing $t$ results in marking $M'$.
Given $\sigma = \langle t_1,t_2,...,t_n \rangle \in T^*$ as a sequence of transitions, $(N,M)[\sigma\rangle(N,M')$ denotes that there is a set of markings $M_0,M_1,...,M_n$ such that $M_0 = M$, $M_n=M'$, and $(N,M_i)[t_{i+1}\rangle(N,M_{i+1})$ for $0 \leq i < n$. $M'$ is called a reachable marking from $M$ if there exists a $\sigma \in T^*$ such that $(N,M)[\sigma\rangle(N,M')$.

\begin{definition}[Labeled Petri Net]
	\label{def:petri_net_labeled}
	A labeled Petri net is a tuple $N = (P,T,F,l)$ where $(P,T,F)$ is a Petri net as defined in Definition~\ref{def:petri_net}, with labeling function $l: T \nrightarrow \universe_{act}$. Given $\sigma_v = \langle a_1,a_2,...,a_n \rangle \in \universe_{act}^*$, $(N,M)[\sigma_v\rangle (N,M')$ if and only if there is a sequence of transitions $\sigma= \langle t_1,t_2,...,t_n \rangle  \in T^*$ such that $(N,M)[\sigma\rangle(N,M')$ and $\sigma_v = l(\sigma)$.
\end{definition}

If $t \notin dom(l)$, it is called an invisible (silent) transition. We write $l(t) = \tau$ if $t \notin dom(l)$. %An occurrence of visible transition $t \in dom(l)$ corresponds to observable activity $l(t) \in \universe_{act}$.
A visible transition $t \in dom(l)$ corresponds to an observable activity $l(t) \in \universe_{act}$.
%In the context of process mining, we always consider processes that start in an initial state and end in a well-defined end state. Thus, we define the notion of a system net.
In the domain of process mining, we always study processes that begin in one state and conclude in another. As a result, we develop the concept of a system net.

\begin{figure}[tb]
    \centering
    \includegraphics[width=0.8\textwidth]{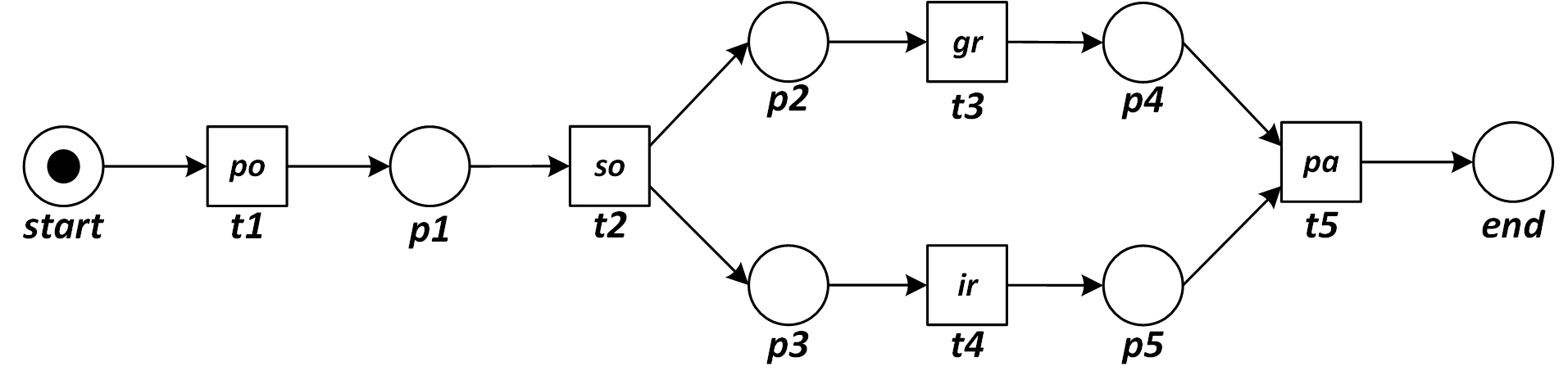}
    \caption{A system net that models the behavior seen in Table~\ref{tbl:sample_event_log}. \\$P{=}\{start,p1,p2,p3,p4,p5,end\},T{=}\{t1,t2,t3,t4,t5\}$, $F{=}\{(start,t1),(t1,p1),(p1,t2),\allowbreak(t2,p2),(t2,p3),(p2,t3),(t3,p4),\allowbreak (p3,t4),\allowbreak(t4,p5),(p4,t5),(p5,t5),(t5,end) \},  M_{init} {=} [start],\\ M_{final}=[end]$.}
    \label{fig:model_sample}
\end{figure}

\begin{definition}[System Net]
	\label{def:system_net}
	A system net is a triplet $SN=(N,M_{init},M_{final})$ where $N=(P,T,F,l)$ is a labeled Petri net, $M_{init} \in \multiset(P)$ is the initial marking, and $M_{final} \in \multiset(P)$ is the final marking. $\universe_{SN}$ denotes the universe of system nets. 
\end{definition}

\begin{definition}[System Net Traces]
	\label{def:system_net_traces}
	Let $SN=(N,M_{init},M_{final})$ be a system net. $\phi_v(SN) = \{ \sigma_v \mid (N,M_{init})[\sigma_v\rangle(N,M_{final}) \}$ is the set of visible traces starting in $M_{init}$ and ending in $M_{final}$.
    $\phi_f(SN) = \{ \sigma \mid (N,M_{init})[\sigma\rangle(N,M_{final}) \}$ is the corresponding set of complete firing sequences.
\end{definition}

Figure~\ref{fig:model_sample} shows a system net that models the behavior seen in Table~\ref{tbl:sample_event_log}, where $\phi_v(SN) = \{ \langle po,so,gr,ir,pa \rangle, \langle po,so,ir,gr,pa \rangle \}$ and $\phi_f(SN) = \{ \langle t1,t2,t3,t4,t5 \rangle,\\ \langle t1,t2,t4,t3,t5 \rangle \}$.

\subsection{Conformance Checking}
Conformance checking techniques investigate how well an event log $L$ and a system net $SN = (N,M_{init},M_{final})$ fit together. Note that $SN$ may have been discovered through process mining or may have been made by hand. In any case, it is interesting to compare the observed example behavior in $L$ with the potential behavior of $SN$. $L$ is perfectly fitting $SN$ if and only if $\bar{L} \subseteq \phi_v(SN)$. 
There are different techniques to quantify fitness such as token-based replay \cite{token_based} and alignments \cite{alignments}. In this paper, we focus on alignments as the most common and well-investigated technique. An alignment between a log and a model is defined based on \textit{moves}.

Given a log $L$ and a model $SN= (N,M_{init},M_{final})$, where $N=(P,T,F,l)$. A \textit{move} refers to a pair $(x,(y,t))$, where the first element $x \in \universe_{act}$ refers to an activity in the log and the second element $(y,t)$ refers to a transition $t \in T$ and its corresponding label $y= l(t)$ in the model. For example, $(a,(a,t1))$ means that both log and model make an \enquote{$a$ move}, a so-called \textit{synchronous move}, where the move in the model is caused by the occurrence of transition $t1$, $(\noMove,(a,t1))$ means that an \enquote{$a$ move} in the model is not mimicked by a corresponding move in the log, a so-called \textit{model move}, and $(a,\noMove)$ means that an \enquote{$a$ move} in the log is not followed by the model, a so-called \textit{log move}.

\begin{definition}[Legal Moves]
	\label{def:legal_moves}
	Let $L \subseteq \univEvents$ be an event log, $A_L = \{ \pi_{act}(e) \mid e \in L\}$ be the set of activities in $L$, and $SN=(N,M_{init},M_{final})$ be a system net, where $N=(P,T,F,l)$. 
    $LM_{SN,L} = \{ (x,(x,t)) \mid x \in A_L \wedge t \in T \wedge l(t)=x \} \cup \{ (\noMove,(x,t)) \mid t \in T \wedge l(t)=x \} \cup \{ (x,\noMove) \mid x \in A_L \}$ is the set of legal moves.
\end{definition}

\begin{definition}[Alignment]
	\label{def:alignment}
	Let $\sigma_L {\in} \tilde{L}$ be a log trace variant and $\sigma_M \in \phi_f(SN)$ be a complete firing sequence of system net $SN$.
    An alignment of $\sigma_L$ and $\sigma_M$ is a sequence of legal moves $\gamma \in LM_{SN,L}^*$, s.t., the projection on the first elements of legal moves (ignoring $\noMove$) results in $\sigma_L$ and the projection on the last element (ignoring $\noMove$ and transition labels) results in $\sigma_M$. 
\end{definition}

For instance, given $\sigma_L= \langle po,gr,ir,so,pa \rangle$ as a log trace and the model shown in Figure~\ref{fig:model_sample}, a corresponding alignment is as follows: 

$$
\gamma = \begin{array}{|c|c|c|c|c|c|}
po & \noMove & gr & ir & so & pa  \\ \hline
po & so & gr & ir & \noMove & pa  \\
t1 & t2 & t3 & t4 &  & t5  \\
\end{array}
$$

Given a trace variant in a log and a process model, there may be many (if not infinitely many) alignments. To select the most appropriate one(s), so-called \textit{optimal alignments}, one needs to associate costs to undesirable moves, i.e., non-synchronous moves (misalignments), and select an alignment with the lowest total costs. 

\begin{definition}[Alignment Cost]
	\label{def:alignment_cost}
	Let $LM_{SN,L}$ be a set of legal moves. A cost function $\delta: LM_{SN,L} \to \natNum$ assigns costs to legal moves. The cost of an alignment $\gamma \in LM_{SN,L}^*$ is the sum of all costs: $\delta(\gamma) = \sum_{(x,(y,t)) \in \gamma}\delta((x,(y,t)))$.
\end{definition}

An alignment $\gamma \in LM_{SN,L}^*$ for a trace variant $\sigma_L {\in} \tilde{L}$ is optimal, if there is no other alignment for the trace whose cost is lower than $\delta(\gamma)$.  
Synchronous moves have no costs, i.e., $\delta(x, (y, t)) = 0$. Model moves only have no costs if the transition is invisible, i.e., $\delta(\noMove, (\tau, t)) = 0$ if $l(t) = \tau$. $\delta(\noMove, (y, t)) > 0$ is the cost of a model move with a visible transition. $\delta(x, \noMove) > 0$ is the cost for a log move. These costs may depend on the nature of the activity, e.g., skipping a payment may be more severe than sending too many letters.
%Note that it is possible to convert misalignment costs into a fitness value between 0 (poor fitness, i.e., maximal costs) and 1 (perfect fitness, zero costs). 
It is worth noting that misaligned costs may be converted into a fitness value ranging from 0 (bad fitness, i.e., maximal costs) to 1 (perfect fitness, zero costs). We refer to \cite{alignments} for details.

\section{Approach}\label{sec:approach}
\begin{figure}[tb]
    \centering
    \includegraphics[width=\textwidth]{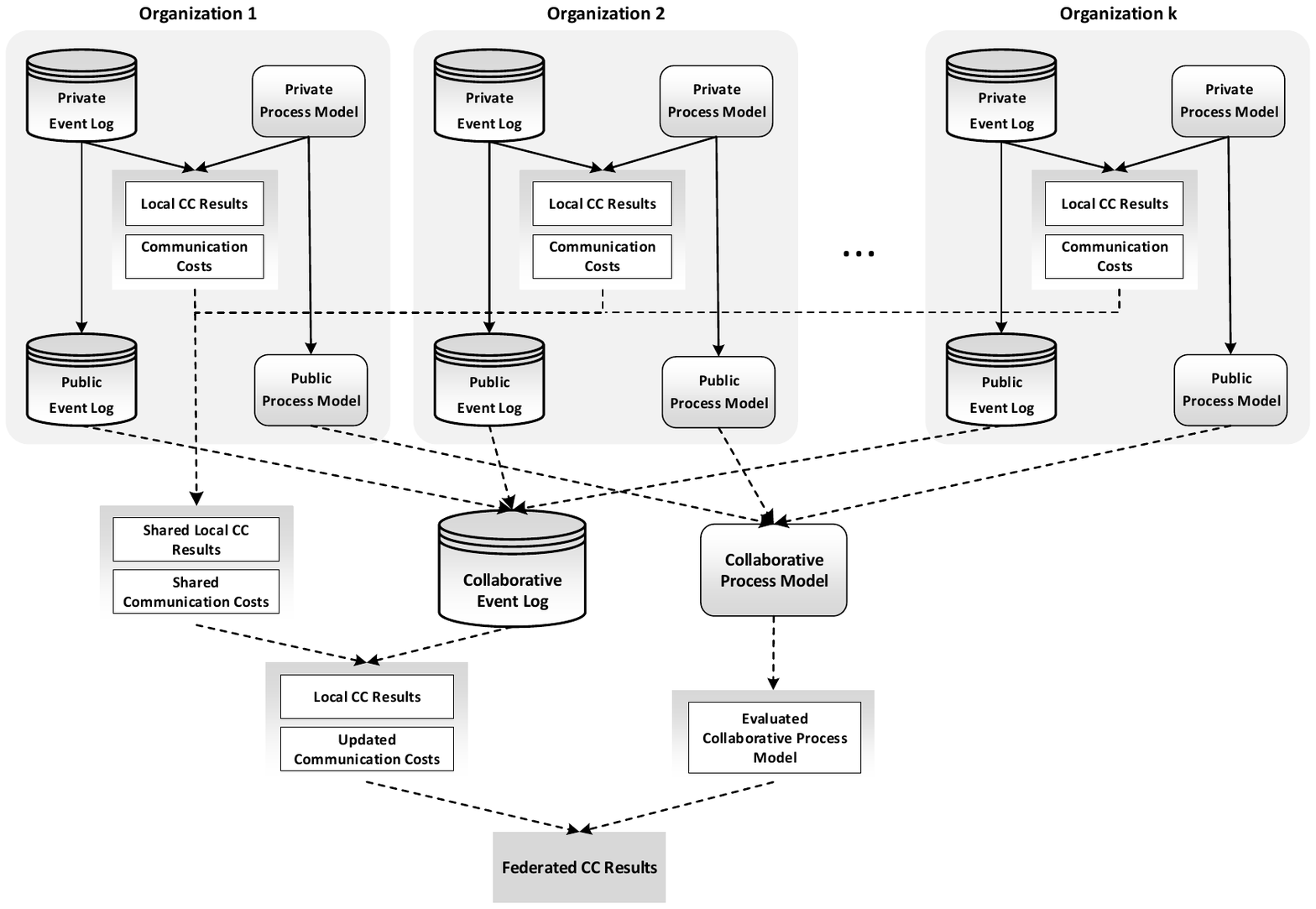}
    \caption{General conceptual framework of federated conformance checking.}
    \label{fig:federated_cc}
\end{figure}

\autoref{fig:federated_cc} depicts the general conceptual framework of federated conformance checking.
A collection of organizations collaboratively execute a process model.
Each organization executes a part of the collaborative process model.
Such an execution emits a data footprint, i.e., a \emph{private event log}.
Moreover, each organization has a private reference model, specifying the intended process behavior.
A private reference model is either created manually or discovered based on the private event log.
Both the private reference process model and the private event log are company-internal and cannot be publicly shared.

\begin{definition}[Private Event Log]
	\label{def:private_event_log}
    Let $\univEvents$ be the universe of events and $\orgs$ be a collection of organizations. A private event log $\prvLog_{o} \subseteq \univEvents$ belonging to an organization $o \in \orgs$ is a collection of events where each event $e {\in} \prvLog_{o}$ has $in$, $out$, and $oid$ as message passing and organization identifier attributes in addition to the default attributes $cid$, $act$, and $time$. 
    % Also, $cid$ contains the organization identifier $O$, i.e., $\pi_{cid}(e) {\in} O \times \universe_{cid}$.
    \vspace{-0.15cm}
    \begin{itemize}
        \item $\pi_{in}(e) \in (\orgs {\setminus} \{o\}) {\cup} \{ \bot \}$ indicates that the event required input from an organization in order to be executed.
        \item $\pi_{out}(e) \in (\orgs {\setminus} \{o\}) \cup \{ \bot \}$ indicates that the execution of the event was required as input from an organization.
        \item $\pi_{oid}(e) \in \orgs$ indicates the identifier of the organization owning the log.
        \item $\{ \pi_{in}(e),\pi_{out}(e) \} \cap \{ \bot\} = \{ \bot \}$. 
        % and $\{ \pi_{in}(e) \} \cup \{ \pi_{out}(e) \} \neq \{ \bot \} $.
    \end{itemize}
\end{definition}

Note the following constraints in Definition~\ref{def:private_event_log} without loss of generality: (1) an organization has no message passing with itself and (2) an event $e$ cannot be executed as a sender, i.e., $\pi_{out}(e) \neq \bot$, and receiver, i.e., $\pi_{in}(e) \neq \bot$, at the same time.
Table~\ref{tbl:private_event_log} shows a private version of the event log shown in Table~\ref{tbl:sample_event_log}, where the manufacturer with identifier $m1$ exchanges messages with a supplier organization with identifier $s1$.

\begin{table}[tb]
\caption{A private version of the sample event log shown in Table~\ref{tbl:sample_event_log}.}
\centering
\scriptsize
\begin{tabular}{l|l|l|l|l|l}
\hline
cid & act             & time & in & out & oid  \\ \hline
c1            & create purchase order ($po$) & 01.01.2023-08:00:00 & $\bot$ & $\bot$ & m1  \\ 
c1            & send order request ($so$)    & 01.01.2023-08:30:00 & $\bot$ & s1 & m1 \\ 
c1            & goods receipt ($gr$)         & 02.01.2023-09:00:00 & $\bot$ & $\bot$ & m1\\ 
c1            & invoice receipt ($ir$)       & 02.01.2023-09:30:00 & s1 & $\bot$ & m1\\ 
c1            & payment ($pa$)               & 02.01.2023-11:00:00 & $\bot$ & s1 & m1\\ 
c2            & create purchase order ($po$) & 02.01.2023-11:30:00 & $\bot$ & $\bot$ & m1\\ 
c2            & send order request ($so$)    & 02.01.2023-11:40:00 & $\bot$ & s1 & m1\\ 
c2            & invoice receipt ($ir$)       & 02.01.2023-16:30:00 & s1 & $\bot$ & m1\\ 
c2            & goods receipt ($gr$)         & 02.01.2023-17:30:00 & $\bot$ & $\bot$ & m1\\ 
c2            & payment ($pa$)               & 03.01.2023-10:30:00 & $\bot$ & s1 & m1\\ 
c3            & create purchase order ($po$) & 03.01.2023-10:40:00 & $\bot$ & $\bot$ & m1\\ 
c3            & send order request ($so$)    & 03.01.2023-10:50:00 & $\bot$ & s1 & m1\\ 
c3            & invoice receipt ($ir$)       & 03.01.2023-16:30:00 & s1 & $\bot$ & m1\\ 
c3            & payment ($pa$)         & 03.01.2023-17:00:00 & $\bot$ & s1 & m1\\ 
c3            & goods receipt ($gr$)               & 03.01.2023-17:30:00 & $\bot$ & $\bot$ & m1\\ \hline
\end{tabular}
\label{tbl:private_event_log}
\end{table}

For a given collection of organizations, $\orgs{=}\{\org_1,\org_2,{\dots}\org_k\}$, we assume that the corresponding private event logs $\prvLog_1,{\dots},\prvLog_k$ do not share any events, i.e., $\forall_{1{\leq}i{<}j{\leq}k}\prvLog_i{\cap}\prvLog_j{=}\emptyset$.
Organizations share a public version of their event log, only consisting of \emph{interaction events}.

\begin{definition}[Public Event Log]
	\label{def:public_event_log}
    Given a private event log of an organization $o$, $\prvLog_{o}$, its corresponding public version is defined as follows: $\pubLog_{o} = \{ e \in \prvLog_{o} \mid \pi_{in}(e) \neq \bot \vee \pi_{out}(e) \neq \bot \}$. Clearly, $\pubLog_{o} \subseteq \prvLog_{o}$.
\end{definition}

Note that private and public event logs can easily be converted to their corresponding simple versions (see Definition~\ref{def:trace}). A \textit{private process model} is a reference model, e.g., a Petri net, that specifies the intended behavior of an organizational process including its interaction with other organizations, i.e., message-passing information modeling an interface. 
We use \textit{open nets}, as a specific type of Petri nets, to define a private process model \cite{open_nets1,behavioral_service}.

\begin{definition}[Private Process Model - Open Net]
	\label{def:private_model}
     Let $I$ be a set of input places and $O$ be a set of output places. $ON^{prv} = (P,T,F,l,M_{init},M_{final},I,O)$ is an open net, modeling a private process model, where:
     \vspace{-0.15cm}
     \begin{itemize}
         \item $(P\cup I\cup O,T,F,l,M_{init},M_{final})$ is a system net such that $P$,$I$,$O$ are pairwise disjoint,
         \item for all $p \in I \cup O$, $M_{init}(p)=0$ and $M_{final}(p)=0$,
         \item for all $p \in I$, $\inputp= \emptyset$, for all $p \in O$, $\outputp= \emptyset$, 
         \item $T_{int} = \{ t \in T \mid (\inputt \cup \outputt) \cap (I \cup O) = \emptyset \}$, $T_{com} = T \setminus T_{int}$, and 
         \item there exists $t {\in} T_{int}$, s.t., $l(t) \neq \tau$.
     \end{itemize}
\end{definition}

\begin{figure}[tb]
    \centering
    \includegraphics[width=0.8\textwidth]{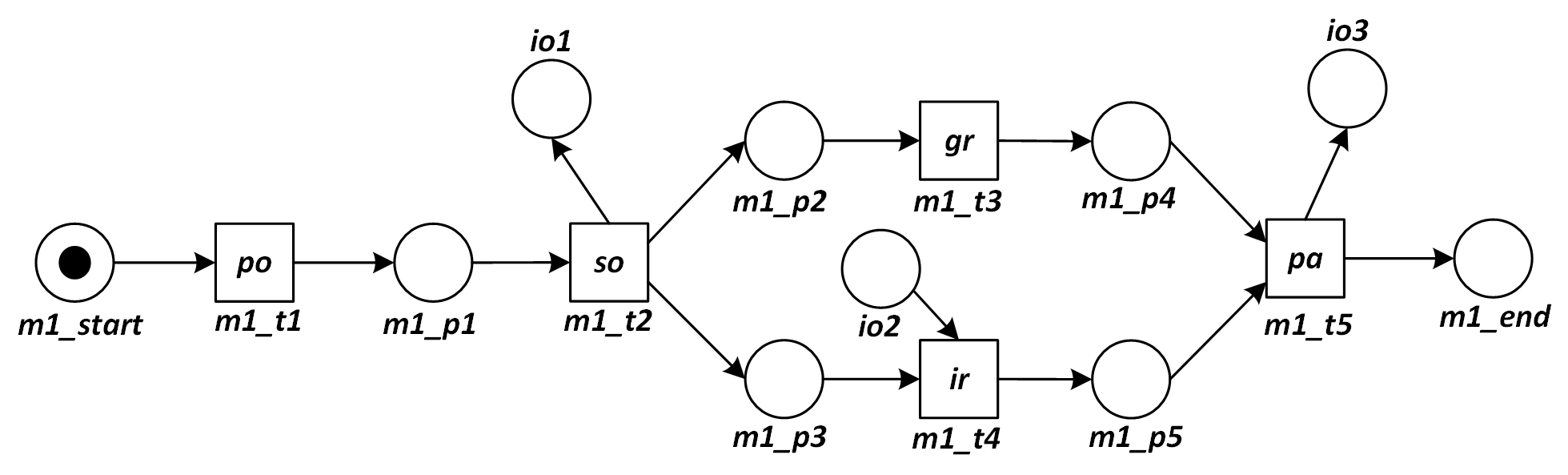}
    \caption{A private process model corresponding to the private event log shown in Table~\ref{tbl:private_event_log}. $I = \{io2\}$, $O = \{ io1,io3 \}$, $T_{int} = \{ m1\_t1,m1\_t3 \}$, and $T_{com} = \{ m1\_t2,m1\_t4,m1\_t5 \}$.}
    \label{fig:private_process_model}
\end{figure}

Note that in Definition~\ref{def:private_model}, $T_{int}$ refers to \textit{internal transitions} that are not involved in any message-passing activities. A set of transitions in a private process model is composed of internal and \textit{communicational transitions} $T_{com}$, i.e., $T = T_{int} \cup T_{com}$. An open net turns to a \textit{closed net} if $I=O=\emptyset$. Given an open net $ON$, $inner(ON)$ is its corresponding system net resulting from removing the interface places and their adjacent arcs from $ON$. Figure~\ref{fig:private_process_model} shows a private process model corresponding to the private event log shown in Table~\ref{tbl:private_event_log}.

Organizations are by default not willing to share their private process models containing their sensitive internal activities. However, we assume that sharing a public version of their private internal process models is of less sensitivity, and organizations are willing to do this to collaboratively analyze the overall collaborative process model.
A public process model only models message-passing activities.  

\begin{definition}[Public Process Model]
	\label{def:public_model}
    The public version of a private process model $ON^{prv} = (P,T,F,l,M_{init},M_{final},I,O)$ is obtained by making all the internal transitions invisible, i.e., $ON^{pub} = (P,T,F,l,M_{init},M_{final},I,O)$ such that for all $t \in T_{int}$, $l(t) = \tau$.   
\end{definition}

\subsection{Local Conformance Checking}\label{subsec:local_CC}
Local conformance checking refers to the alignment of a private event long (i.e., observed behavior) and a private process model (i.e., modeled behavior) in each organization. In order to align observed behavior and modeled behavior, we need as input a simple event log $L \in \multiset(\universe_{act}^*)$ and a
system net $SN = (N,M_{init},M_{final})$. A simple event log can easily be obtained from a private event log (see Definition~\ref{def:trace}). A system net of a private process model $ON^{prv}$ is obtained by $inner(ON^{prv})$.   
Note that private process models represented by open nets cannot simply be taken as input for conformance checking. That is because in an open net with input places $I$ and output places $O$, transitions consuming from $I$ are dead, and tokens produced on places in $O$ cannot be removed by the net.

\subsection{Local Communication Costs}\label{subsec:local_com_cost}
To compute federated conformance checking, each organization needs to calculate the local communication cost associated with a communication point of its private process model (i.e., an input or an output place). The local communication cost of a communication point refers to the \textit{potential} cost of misalignment between a trace and a private process model (i.e., an open net) due to a miscommunication between two organizations. 
We introduce three different types of miscommunications: \textit{sender move}, \textit{receiver move}, and \textit{asynchronous communication}. In the following, we define different types of (mis)communications.

\begin{definition}[Communication Types]
	\label{def:com_types}
    Let $o_s$ and $o_r$ be sender and receiver organizations communicating through a place $p_1$ used as output place in the process model of $o_s$ and as input place in the process model of $o_r$.
    A \textbf{sender move} refers to a situation where a message (i.e., a token in a Petri net) produced by $o_s$ in $p_1$ is never consumed by $o_r$. 
    A \textbf{receiver move} refers to a situation where $o_r$ consumes a message from $p_1$ that is never produced by $o_s$. 
    An \textbf{asynchronous communication} refers to a situation where $o_r$ consumes a message from $p_1$ before getting produced by $o_s$.
\end{definition}

Note that miscommunications are not observable in an organization without having access to the communication-related data of organizations involved in communications. Thus, local communication costs are considered potential costs and calculated by assuming that traces involved in communications misalign with the organization's private process model due to miscommunications. 
To calculate local communication costs (i.e., potential miscommunications), we replace the label of communicational transitions on the private model with a label that cannot get synched with the corresponding activity in the trace, e.g., the silent label $\tau$.

We consider two specific types of labels that can be replaced with communicational transitions: input labels and output labels.

\begin{itemize}
    \item \textbf{Input labels}: These are associated with transitions that connect to an input place in the process model. We replace an input label with the label of a communicational transition that is linked to this input place.
    \item \textbf{Output labels}: These are tied to transitions that connect to an output place. Similarly, an output label is replaced with the label of a communicational transition associated with this output place.
\end{itemize}
   
In practice, different misalignment costs can be assigned to input labels and output labels, depending on the severity or significance of the misalignment in the context of the process being analyzed.
To represent these labels, we use the labels $\tau_{in}$ for input labels and $\tau_{in}$ for output labels. We treat these labels as silent transitions. This means that when these labels are used, they do not incur any cost in terms of a corresponding model move.
However, it is important to note that when the label of a communicational transition is replaced with specific silent labels ($\tau_{in}$ or $\tau_{out}$), it leads to a log move in the event logs. This log move reflects the occurrence of an event in the trace without a corresponding action in the model, effectively indicating a communication-related discrepancy. The minimal cost that can be assigned to this type of miscommunication is based on this log move, representing the least penalty for such a misalignment.

Moreover, as discussed in Subsection~\ref{subsec:local_CC}, alignments cannot be computed over open nets. Thus, after replacing the label of the communicational transition associated with a communication point, for which we want to calculate the local communication cost, we obtain the corresponding system net by applying the inner function $inner(.)$ to the open net. Note that the local communication cost of a trace includes the local alignment cost of the trace. Thus, we need to subtract the local conformance cost from the local communication cost to obtain the cost solely associated with potential miscommunications.

\begin{figure}[]
    \centering
    \includegraphics[width=0.9\textwidth]{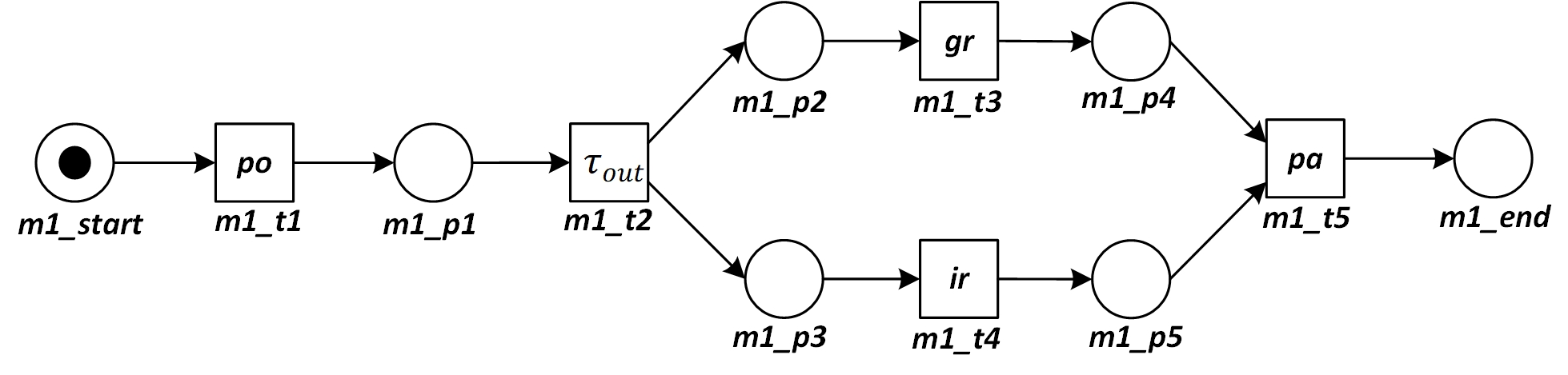}
    \caption{The system net obtained from Figure~\ref{fig:private_process_model} by replacing the label of communication transition $m1\_t2$ with $\tau_{out}$ as an output label and applying the $inner(.)$ function.}
    \label{fig:private_process_model_inner}
\end{figure}

As an example, consider a trace $\sigma = \langle po, so, ir, pa, gr \rangle$ and the private process model in Figure~\ref{fig:private_process_model}. To calculate the local communication cost associated with communication point $io1$, we replace the label of communication transition $m1\_t2$ with $\tau_{out}$ as an output label and apply function $inner(.)$ to the open net. Figure~\ref{fig:private_process_model_inner} shows the corresponding system net. The alignment of trace $\sigma$ with the resulting system net is as follows:

$$
\footnotesize
\gamma = \begin{array}{|c|c|c|c|c|c|c|}
po & so & \noMove & ir & pa & gr & \noMove  \\ \hline
po & \noMove & \tau_{out} & ir & \noMove & gr & pa  \\
m1\_t1 &  & m1\_t2 & m1\_t4 &  & m1\_t3 & m1\_t5  \\
\end{array}
$$

As can be seen in the alignment of the above-mentioned example, the local communication cost associated with communication point $io1$ appears as a log move associated with activity $so$ that cannot be mimicked by the model because of the replacement of the corresponding activity label with a silent transition. Considering no cost for the output label $\tau_{out}$ the local communication cost is 1. As mentioned earlier, one can consider different costs for output (input) labels. However, the overall cost of this alignment, which is 3 assuming no cost for the output label, includes the local alignment cost as well, which is due to the order misalignment of activities $pa$ and $gr$. To obtain the local communication cost associated with communication point $io1$ excluding the local alignment cost, we need to subtract the local alignment cost from the overall alignment cost. Thus, the local communication cost associated with $io1$ is $3-2=1$.

In the remainder, the local communication cost of a trace for a specific communication point refers to the cost that does not contain the local conformance cost of the trace.

\subsection{Communication Between Public Process Models}
Communication between two public process models is modeled by composing the respective open nets \cite{behavioral_service}. To this end, shared input and output places are merged and turned into internal places. %A merged internal place models a channel, and a token on such a place corresponds to a pending message in the channel. 
A channel is represented by a merged internal place, and a token on such a place corresponds to a waiting message in the channel. For the composition of two open nets, we assume that the sets of transitions and internal places of two nets are pairwise disjoint. However, the interfaces, i.e., the sets of input and output places, overlap.
% \footnote{\scriptsize This can be done in practice by considering a set that contains the sender and receiver organization ids as the identifier of their interface place, e.g., $\{o_s,o_r\}$ can be the identifier of an interface place between a sender $o_s$ and a receiver $o_r$.} 
%Moreover, every shared place $p$ has only one open net that sends into $p$ and one open net that receives from $p$,
Furthermore, each shared place $p$ has just one open net that sends into $p$ and one open net that receives from $p$. i.e., communication is \textit{bilateral} and \textit{directed} \cite{behavioral_service}. Open nets that meet these requirements are called \textit{composable}.

\subsection{Running Example}
\begin{figure}[tb]
    \centering
    \includegraphics[width=0.9\textwidth]{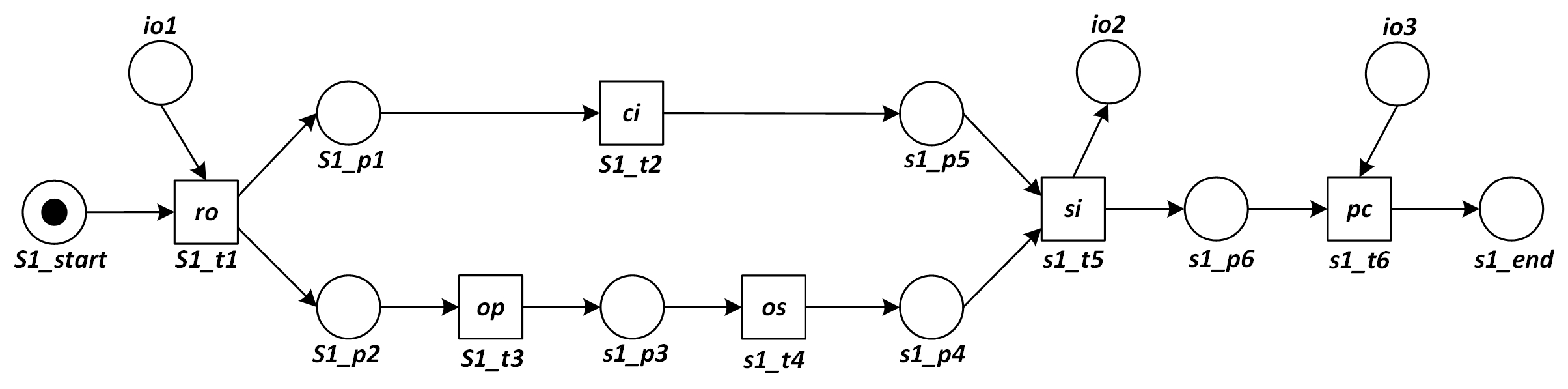}
    \caption{A private process model corresponding to the private event log shown in Table~\ref{tbl:private_event_log_supplier}. $I = \{io1,io3\}$, $O = \{ io2 \}$, $T_{int} = \{ s1\_t2,s1\_t3,s1\_t4 \}$, and $T_{com} = \{ s1\_t1,s1\_t5,s1\_t6 \}$.}
    \label{fig:private_process_model_supplier}
\end{figure}

As a running example, we consider a simplified supply chain process, where a collaborative process is executed by two parties: a \emph{manufacturer}, executing a purchase-to-pay process, and a \textit{supplier} executing an order-to-cash process.
We consider the process model shown in Figure~\ref{fig:private_process_model} as the private process model of the manufacturer and the process model shown in Figure~\ref{fig:private_process_model_supplier} as the private process model of the supplier, with a corresponding private event log shown in Table~\ref{tbl:private_event_log_supplier}.

\begin{table}[tb]
\caption{A fragment of a private event log recorded by a supplier's information system by executing an order-to-cash process.}
\centering
\scriptsize
\begin{tabular}{l|l|l|l|l|l}
\hline
cid & act             & time & in & out & oid  \\ \hline
c1            & receive order ($ro$) & 01.01.2023-08:40:00 & m1 & $\bot$ & s1  \\ 
c1            & create invoice ($ci$)    & 01.01.2023-08:50:00 & $\bot$ & $\bot$ & s1 \\ 
c1            & order processing ($op$)         & 01.01.2023-09:00:00 & $\bot$ & $\bot$ & s1\\ 
c1            & order shipment ($os$)       & 01.01.2023-15:30:00 & $\bot$ & $\bot$ & s1\\ 
c1            & send invoice ($si$)               & 02.01.2023-09:00:00 & $\bot$ & m1 & s1\\ 
c1            & payment collection ($pc$) & 02.01.2023-11:30:00 & m1 & $\bot$ & s1\\ 
c2            & receive order ($ro$)    & 02.01.2023-11:50:00 & m1 & $\bot$ & s1\\ 
c2            & order processing ($op$)       & 02.01.2023-12:30:00 & $\bot$ & $\bot$ & s1\\ 
c2            & order shipment ($os$)         & 02.01.2023-16:30:00 & $\bot$ & $\bot$ & s1\\ 
c2            & create invoice ($ci$)               & 02.01.2023-17:30:00 & $\bot$ & $\bot$ & s1\\ 
c2            & send invoice ($si$)               & 02.01.2023-17:40:00 & $\bot$ & m1 & s1\\
c2            & payment collection ($pc$)               & 03.01.2023-10:40:00 & m1 & $\bot$ & s1\\
c3            & receive order ($ro$)    & 03.01.2023-10:55:00 & m1 & $\bot$ & s1\\ 
c3            & create invoice ($ci$)               & 03.01.2023-11:00:00 & $\bot$ & $\bot$ & s1\\ 
c3            & order processing ($op$)       & 03.01.2023-11:30:00 & $\bot$ & $\bot$ & s1\\ 
c3            & send invoice ($si$)               & 03.01.2023-16:20:00 & $\bot$ & m1 & s1\\
c3            & order shipment ($os$)         & 03.01.2023-17:00:00 & $\bot$ & $\bot$ & s1\\ 
\hline
\end{tabular}
\label{tbl:private_event_log_supplier}
\end{table}

\begin{definition}[Collaborative Process Model - Open Nets Composition]
	\label{def:collab_model}
    Let $ON^{pub}_1 = (P_1,T_1,F_1,l_1,M_{{init}_1},M_{{final}_1},I_1,O_1)$ and $ON^{pub}_2 = (P_2,T_2,\\F_2,l_2,M_{{init}_2},M_{{final}_2},I_2,O_2)$ be two open nets modeling two public process models. $ON^{pub}_1$ and $ON^{pub}_2$ are composable if $(P_1\cup T_1 \cup I_1 \cup O_1) \cap (P_2\cup T_2 \cup I_2 \cup O_2) = (I_1 \cap O_2) \cup (I_2 \cap O_1)$. The composition of $ON^{pub}_1$ and $ON^{pub}_2$ is an open net $ON^{pub}_1 \bigoplus ON^{pub}_2 = (P,T,F,l,M_{init},M_{final},I,O)$ modeling a collaborative process model where:
    \vspace{-0.15 cm}
    \begin{itemize}
        \item $P = P_1 \cup P_2 \cup (I_1 \cap O_2) \cup (I_2 \cap O_1)$,
        \item $T = T_1 {\cup} T_2$, $F=F_1 {\cup} F_2$, $M_{init} = M_{init_1} {\uplus} M_{init_2}$, $M_{final} = M_{final_1} {\uplus} M_{final_2}$,
        \item $l: T \nrightarrow \universe_{act}$ with $dom(l) = dom(l_1) \cup dom(l_2)$, $l(t) = l_1(t)$ if $t \in dom(l_1)$ and $l(t) = l_2(t)$ if $t \in dom(l_2)$, 
        \item $I = (I_1 \cup I_2) \setminus (O_1 \cup O_2)$, and $O = (O_1 \cup O_2) \setminus (I_1 \cup I_2)$.
    \end{itemize}
    
\end{definition}

\begin{figure}[tb]
    \centering
    \includegraphics[width=0.95\textwidth]{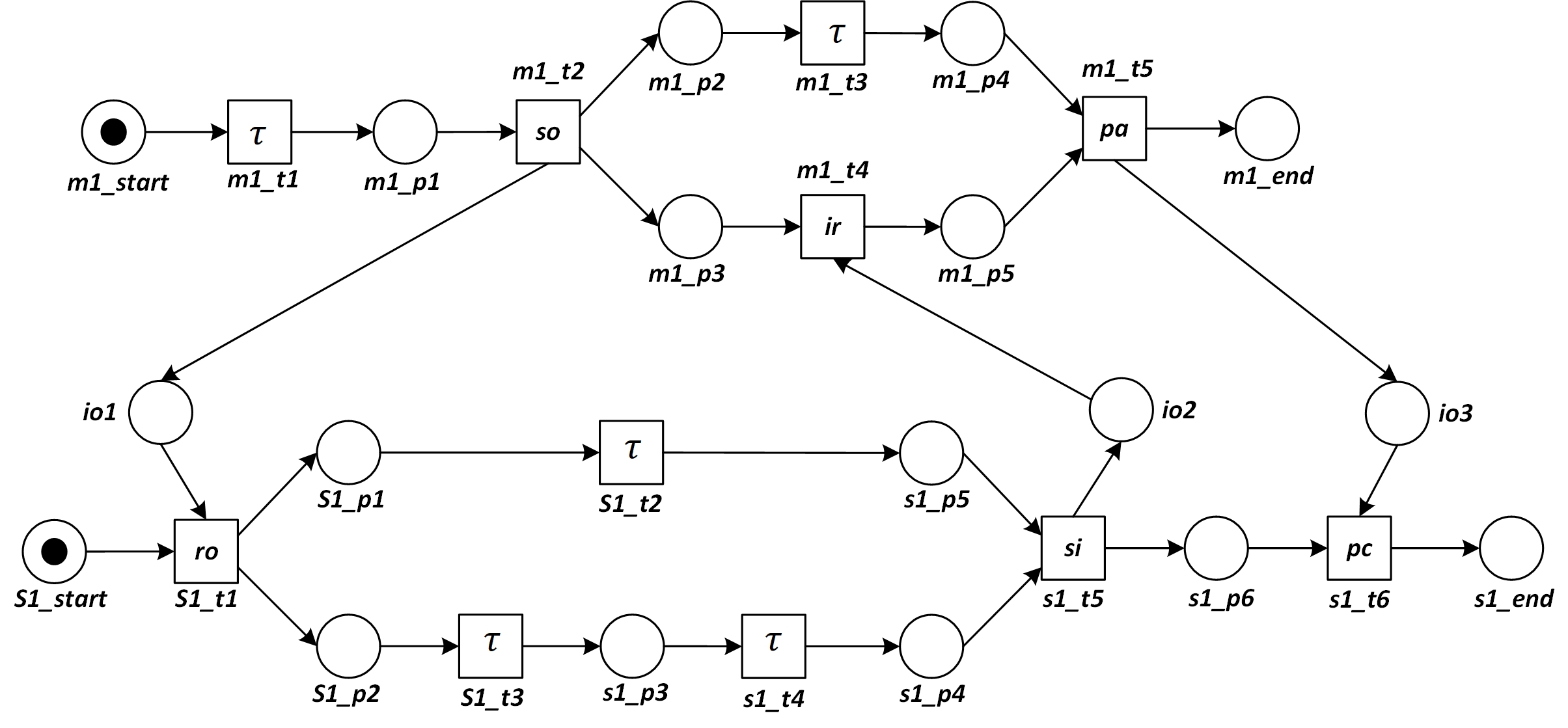}
    \caption{The collaborative model obtained by composing the public versions of the private process models shown in Figure~\ref{fig:private_process_model} and Figure~\ref{fig:private_process_model_supplier}.}
    \label{fig:collaborative_model}
\end{figure}

Note that open net composition models \textit{asynchronous} message-passing. Asynchronous message-passing implies non-blocking communication, where a process can proceed with its execution without waiting for a sent message to be received. Additionally, messages can be overtaken by one another, meaning that the order in which messages are sent may not align with the order in which they are received.
Figure~\ref{fig:collaborative_model} shows the collaborative model obtained by composing the public versions of the private process models shown in Figure~\ref{fig:private_process_model} and Figure~\ref{fig:private_process_model_supplier}.

We want a collaborative process model to be \textit{valid}. A valid collaborative process model is a closed net, where $I=O=\emptyset$. That is, after composing public process models there must not be input or output places remaining unmerged, i.e., not turned into internal places of the collaborative process model. The validity checking of a collaborative process model is the first step of federated conformance checking that can disclose communicational problems.
Assuming that the collaborative process model is a valid closed net, we can compute federated conformance checking based on shared local conformance checking results, shared local communication costs, and the collaborative event log which is a collection of public event logs.

\begin{definition}[Collaborative Event Log]
	\label{def:collab_EL}
    Given $\pubLog_1, \pubLog_2, ..., \pubLog_k$ as public logs of $k$ organizations, $ L^{col} = \pubLog_1 \cup \pubLog_2 \cup ... \cup \pubLog_k$ is their corresponding collaborative event log. Given $L^{col}$, the public event log of an organization $o$ is $\pubLog_{o} = \{ e \in L^{col} \mid \pi_{oid}(e) = o \}$.  
\end{definition}

Table~\ref{tbl:collaborative_log} shows the collaborative log obtained by combining the public versions of the private event logs shown in Table~\ref{tbl:private_event_log} and Table~\ref{tbl:private_event_log_supplier}. 

The first step of calculating federated alignment costs is to use the collaborative event log to realize (real) communication costs of local communication costs calculated by each organization. As mentioned in Subsection~\ref{subsec:local_com_cost}, a local communication cost of a trace refers to the cost associated with a potential miscommunication related to a communication point.

\begin{table}[tb]
\caption{The collaborative log obtained by combining the public versions of the private event logs, shown in Table~\ref{tbl:private_event_log} and Table~\ref{tbl:private_event_log_supplier}.}
\centering
\scriptsize
\begin{tabular}{l|l|l|l|l|l}
\hline
cid & act             & time & in & out & oid  \\ \hline
c1            & send order request ($so$)    & 01.01.2023-08:30:00 & $\bot$ & s1 & m1 \\ 
c1            & receive order ($ro$) & 01.01.2023-08:40:00 & m1 & $\bot$ & s1  \\ 
c1            & send invoice ($si$)               & 02.01.2023-09:00:00 & $\bot$ & m1 & s1\\ 
c1            & invoice receipt ($ir$)       & 02.01.2023-09:30:00 & s1 & $\bot$ & m1\\ 
c1            & payment ($pa$)               & 02.01.2023-11:00:00 & $\bot$ & s1 & m1\\ 
c1            & payment collection ($pc$) & 02.01.2023-11:30:00 & m1 & $\bot$ & s1\\ 
c2            & send order request ($so$)    & 02.01.2023-11:40:00 & $\bot$ & s1 & m1\\ 
c2            & receive order ($ro$)    & 02.01.2023-11:50:00 & m1 & $\bot$ & s1\\ 
c2            & send invoice ($si$)               & 02.01.2023-17:40:00 & $\bot$ & m1 & s1\\
c2            & invoice receipt ($ir$)       & 02.01.2023-18:30:00 & s1 & $\bot$ & m1\\ 
c2            & payment ($pa$)               & 03.01.2023-10:30:00 & $\bot$ & s1 & m1\\ 
c2            & payment collection ($pc$)               & 03.01.2023-10:40:00 & m1 & $\bot$ & s1 \\
c3            & send order request ($so$)    & 03.01.2023-10:50:00 & $\bot$ & s1 & m1\\
c3            & receive order ($ro$)    & 03.01.2023-10:55:00 & m1 & $\bot$ & s1\\ 
c3            & send invoice ($si$)               & 03.01.2023-16:20:00 & $\bot$ & m1 & s1\\
c3            & invoice receipt ($ir$)       & 03.01.2023-16:30:00 & s1 & $\bot$ & m1\\ 
c3            & payment ($pa$)         & 03.01.2023-17:00:00 & $\bot$ & s1 & m1\\ 
\hline
\end{tabular}
\label{tbl:collaborative_log}
\end{table}

\begin{table}[b]
\scriptsize
\caption{The local alignment and communication costs for trace $\langle po, so, ir, pa, gr \rangle$ with respect to the system net of the private model of organization m1 shown in Figure~\ref{fig:private_process_model}, considering no cost for input and output labels.}
\centering
\begin{tabular}{|c|c|c|c|c|}
\hline
case id & \begin{tabular}[c]{@{}l@{}}local align. cost \\ $LaC^{m_1}(c3)$\end{tabular} & \begin{tabular}[c]{@{}l@{}}local com. cost io1 \\ $LcC^{m_1}(io1)$\end{tabular} & \begin{tabular}[c]{@{}l@{}}local com. cost io2 \\ $LcC^{m_1}(io2)$\end{tabular} & \begin{tabular}[c]{@{}l@{}}local com. cost io3 \\ $LcC^{m_1}(io3)$ \end{tabular} \\ \hline
c3      & 2                                                            & 1                                                              & 1                                                             & 1                                                             \\ \hline
\end{tabular}
\label{tbl:cc_results_m1}
\end{table}

\begin{table}[tb]
\scriptsize
\caption{The local alignment and communication costs for trace $\langle ro, ci, op, si, os \rangle$ with respect to the system net of the private model of organization s1 shown in Figure~\ref{fig:private_process_model_supplier}, considering no cost for input and output labels.}
\centering
\begin{tabular}{|c|c|c|c|c|}
\hline
case id & \begin{tabular}[c]{@{}l@{}}local align. cost \\ $LaC^{s_1}(c3)$\end{tabular} & \begin{tabular}[c]{@{}l@{}}local com. cost io1 \\ $LcC^{s_1}(io1)$\end{tabular} & \begin{tabular}[c]{@{}l@{}}local com. cost io2 \\ $LcC^{s_1}(io2)$\end{tabular} & \begin{tabular}[c]{@{}l@{}}local com. cost io3 \\ $LcC^{s_1}(io3)$\end{tabular} \\ \hline
c3      & 3                                                            & 1                                                              & 1                                                             & 0                                                             \\ \hline
\end{tabular}
\label{tbl:cc_results_s1}
\end{table}

The local communication cost of a trace $\sigma_c$ belonging to a case $c$ and related to a communication point $io$ becomes a real communication cost if the type of communication associated with $io$ \textit{is not synchronous}. We consider \textbf{no cost} for a local communication cost if it is not realized after sharing public event logs. The federated alignment cost of a case involved in communication between two organizations is calculated as Definition~\ref{def:federated_align_cost}. 

\begin{definition}[Federated Alignment Cost]
	\label{def:federated_align_cost}
   Let $org_1$ and $org_2$ be two organizations involved in communication via a joint case $c$. Given $LaC^{org_1}(c)$ as the local alignment cost of $c$ in $org_1$, $LaC^{org_2}(c)$ as the local alignment cost of $c$ in $org_2$, $LcC^{org_1}(c)$ as the real communication cost of $c$ in $org_1$ for all the communication points, and $LcC^{org_2}(c)$ as the real communication cost of $c$ in $org_2$ for all the communication points, the federated alignment cost of case $c$ is $FaC(c) = LaC^{org_1}(c) + LaC^{org_2}(c) + LcC^{org_1}(c) + LcC^{org_2}(c)$.   
\end{definition}

As an example, consider $\sigma_{c3}^{m1} =  \langle po, so, ir, pa, gr \rangle$ and  $\sigma_{c3}^{s1} =  \langle ro, ci, op, si, os \rangle$ as the trace of case $c3$ in organizations $m1$ and $s1$, respectively. Tables~\ref{tbl:cc_results_m1} and \ref{tbl:cc_results_s1} show the local alignment and communication costs with respect to the system nets of the private models of two organizations shown in Figures~\ref{fig:private_process_model} and \ref{fig:private_process_model_supplier}, considering no cost for input and output labels.

The alignment of the collaborative trace $\sigma_{c3}^{col} = \langle so,ro,si,ir,pa \rangle$, which can be obtained from the collaborative event log with the collaborative process model is as follows:

$$
\scriptsize
\gamma = \begin{array}{|c|c|c|c|c|c|c|c|c|c|c|c|}
\noMove & so & ro & \noMove & \noMove & \noMove & si & ir & \noMove & pa & \noMove \\ \hline
\tau & so & ro & \tau & \tau & \tau & si & ir & \tau & pa & pc \\
m1\_t1 & m1\_t2 & s1\_t1 & s1\_t2 & s1\_t3 & s1\_t4 & s1\_t5 & m1\_t4 & m1\_t3 & m1\_t5 & s1\_t6 \\
\end{array}
$$

In the alignment of a collaborative trace with the collaborative process model, each visible non-synchronous move (see Definition~\ref{def:alignment} and different types of moves) is associated with a type of miscommunication. In the above alignment, there exists only one visible non-synchronous move, which is associated with a sender move type of miscommunication related to the communication point $io3$. Thus, the only local miscommunication that is realized is the one associated with the communication point $io3$. The federated alignment cost of case $c3$ is calculated as follows: $LaC^{m1}(c3) = 2$, $LaC^{s1}(c3) = 3$, $LcC^{m1}(c3) = 0+0+1$, $LcC^{s1}(c3) = 0+0+0$, and $FaC(c3)= 3+2+1+0$.   

Note that assuming that miscommunication costs are not customized by individual organizations, the federated alignment cost for a case is equal to the local alignment costs of each organization plus the alignment costs of the corresponding collaborative trace with respect to the collaborative process model.

\subsection{Privacy Considerations}
In general, two main levels for privacy concerns can be considered in federated process mining: \textit{individual-level} and \textit{organizational-level} \cite{rafiei_federated_process}. 
The former aims to protect private data belonging to individuals in organizations.
The latter considers the sensitive internal activities of an organization as private information that should not be revealed \cite{liu2019towards,rafiei_federated_process}. As described below, our approach considers both levels of privacy concerns. 

Focusing on the control-flow aspect of event logs, the sensitive individual-related data that need to be protected are complete sequences of activities performed for individuals (cases) \cite{rafiei_cedp_dp,gamal_libra,stephan_sacofa,rafiei_travas,rafiei_travag}.
In our approach, we never share a complete sequence of activities performed for a case. We only share the local alignment results per case and public models that cannot be exploited to infer complete sequences of activities.

The sensitive organizational-level data that need to be protected, focusing on the control-flow aspect of event logs, are internal organization-specific activities that are never shared in our approach. We only share communicational activities that are involved in sending (receiving) messages to (from) other organizations.

\section{Experimental Results}\label{sec:evaluation}
In order to validate the effectiveness of our proposed approach, we devised a simulated supply chain process involving three organizations: a manufacturer, a supplier, and a shipper. Specifically, we focused on purchase-to-pay (executed by the manufacturer), order-to-cash (executed by the supplier), and shipment (executed by the shipper) processes, which are critical components of the supply chain domain. 

\begin{figure}[htbp]
    \centering
    \includegraphics[width=.9\textwidth,keepaspectratio]{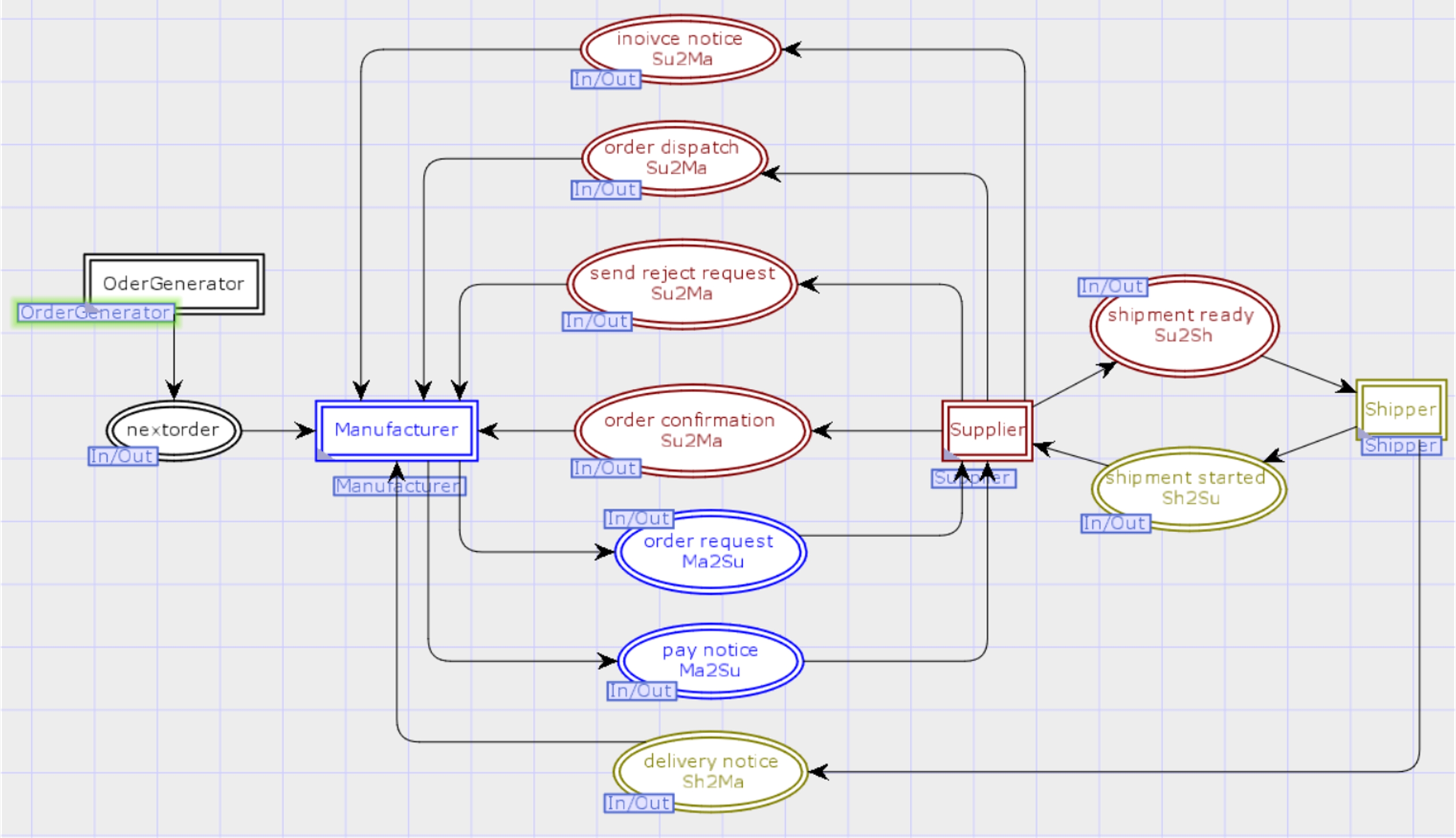}
    \caption{The high-level designed process and communications among three organizations in the CPN Tools. }
    \label{fig:EvaluationTop}
\end{figure}

\begin{figure}[htbp]
    \centering
    %ManufacturerVisio.pdf
    \includegraphics[height=\textheight,keepaspectratio]{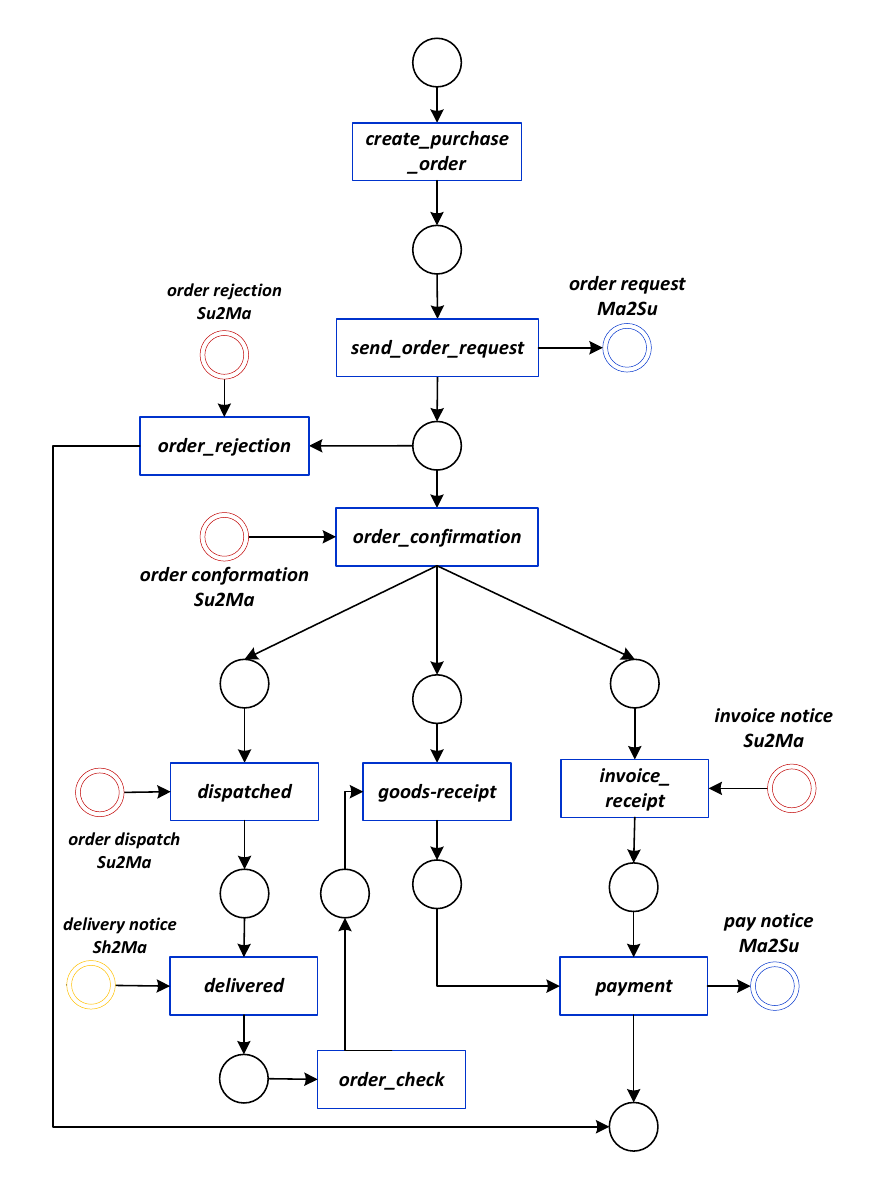}
    \caption{The designed and implemented process of the manufacturer.}% in the CPN Tools. }
    \label{fig:EvaluationsManu}
\end{figure}

We used the CPN Tools \cite{CPNTools2003} and SML functions to simulate this scenario. The source code is available in our GitHub repository\footnote{\scriptsize \href{https://github.com/m4jidRafiei/Federated\_Conformance\_Checking/tree/main/CPN}{https://github.com/m4jidRafiei/Federated\_Conformance\_Checking/tree/main/CPN}}. 
The abstract high-level design and communication among three organizations in the explained supply chain process is shown in Figure~\ref{fig:EvaluationTop}. Figure~\ref{fig:EvaluationsManu} represents the detailed process inside the manufacturer, where it starts with the creation of a purchase order and is followed by the order confirmation or rejection. This process ends with payment. The designed processes of the supplier and shipper are shown in Figure~\ref{fig:SupplierShipper3}.
By designing this synthetic supply chain process, we have full control over its characteristics, allowing us to evaluate our approach in a controlled environment. 

\begin{sidewaysfigure}[htbp]
    \centering
    %SupplierVisio.pdf
    \includegraphics[width=\textwidth,keepaspectratio]{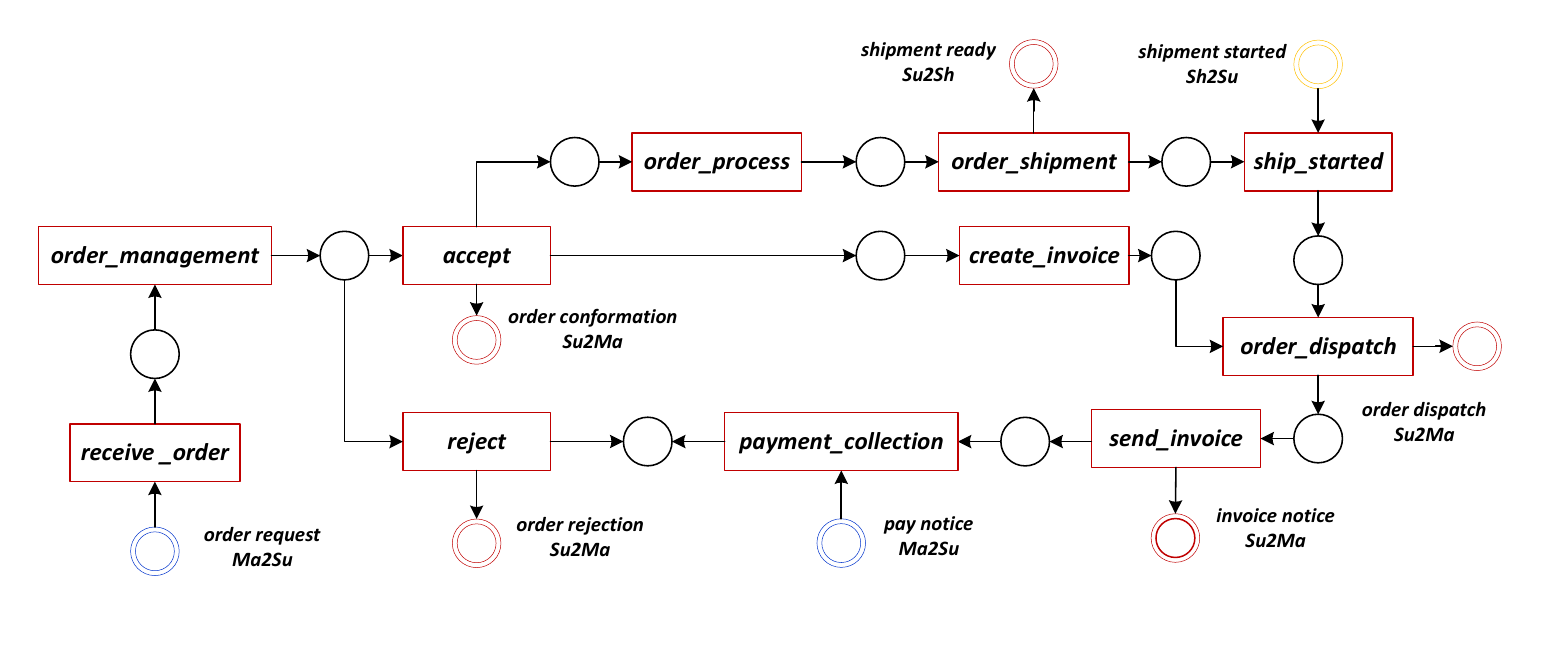}
    \caption{The designed and implemented processes of the supplier.}
    \label{fig:SupplierShipper3}
\end{sidewaysfigure}

\begin{figure}[htbp]
    \centering
    \includegraphics[width=1.05\textwidth,keepaspectratio]{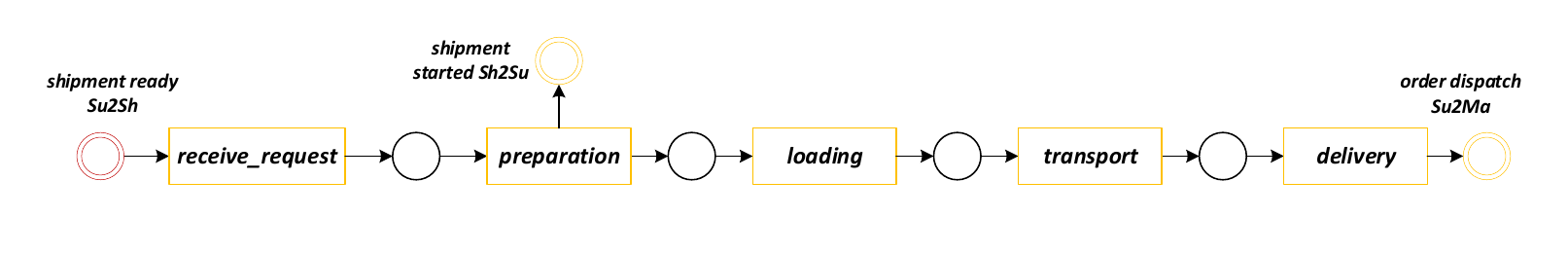}
    \caption{The designed and implemented processes of the shipper.}
    \label{fig:SupplierShipper3}
\end{figure}

\begin{comment}
\begin{figure}[t]
    \centering
    \includegraphics[width=.9\textwidth,keepaspectratio]{figs/SupplierShipper3.jpg}
    \caption{The designed and implemented processes of the supplier and shipper in the CPN Tools.}
    \label{fig:SupplierShipper3}
\end{figure}
\end{comment}

Using the designed process models in the CPN Tools, we generate synthetic event logs capturing the activities and interactions within the supply chain process. These event logs served as the basis for our conformance checking analysis. The generated event logs are also available in the GitHub repository\footnote{\scriptsize \href{https://github.com/m4jidRafiei/Federated\_Conformance\_Checking/tree/main/Logs}{https://github.com/m4jidRafiei/Federated\_Conformance\_Checking/tree/main/Logs}}. Table~\ref{tbl:logs_stat} shows the general statistics of the event logs. 
% By applying our proposed approach to these synthetic event logs, we were able to assess the accuracy of our method in detecting and quantifying costs associated with pre-injected miscommunications.

\begin{table}[!htb]
\small
\caption{The general statistics of the generated event logs.}
\centering
\begin{tabular}{|c|c|c|c|c|}
\hline
            & \#cases & \#events & \#activities & \#communication points \\ \hline
Manufacture & 297     & 2517     & 10           & 7                      \\ \hline
Supplier    & 297     & 2788     & 11           & 8                      \\ \hline
Shipper     & 271     & 1355     & 5            & 3                      \\ \hline
Overall     & 297     & 6660     & 26           & 9                      \\ \hline
\end{tabular}
\label{tbl:logs_stat}
\end{table}

By deliberately injecting miscommunications into the synthetic event logs, we introduced various scenarios representing communication breakdowns or deviations from the expected process flow. Our approach then enabled us to identify the costs incurred due to these miscommunications. This analysis provides valuable insights into the impact of miscommunications on the overall process performance and allows us to evaluate the effectiveness of our approach in capturing such costs.

\subsection{Miscommunication Scenarios}
We inject three miscommunication scenarios corresponding to the three miscommunication types including \textit{sender move}, \textit{receiver move}, and \textit{asynchronous communication}.

\begin{itemize}
    \item \textbf{Sender move}: To simulate this type of miscommunication, we focus on the communication between the manufacturer and the supplier through the communication point \textit{order dispatch}. A sender move happens when the supplier as the sender executes the order dispatch activity but the manufacturer does not execute the buddy activity that is dispatched. Thus, we deliberately removed the dispatched activity from 10 cases in the event log of the manufacturer.
    \item \textbf{Receiver move}: To simulate this type of miscommunication, we focus on the communication between the supplier and the shipper via the communication point \textit{shipment started}. A receiver move happens when the supplier executes the activity ship\_started while the shipper does not execute the buddy activity which is preparation. We intentionally removed the preparation activity from the shipper's event log for 19 cases.
    \item \textbf{Asynchronous communication}: This type of miscommunication refers to the situation where the communication happens in the wrong order. We focus on the communication between the shipper and the manufacturer to simulate this type of communication. Particularly, we swap the order of the \textit{delivery} and \textit{delivered} activities that are connected via the communication point delivery notice such that the delivered activity is executed by the manufacturer earlier before the delivery activity is executed by the shipper. To this end, we shifted the execution time of the delivery activity by the shipper for 5 cases.   
\end{itemize}

The altered event logs with the inserted miscommunications for three organizations are available in the GitHub repository\footnote{\scriptsize \href{https://github.com/m4jidRafiei/Federated\_Conformance\_Checking/tree/main/Modified Logs}{https://github.com/m4jidRafiei/Federated\_Conformance\_Checking/tree/main/Modified Logs}}
. 

\begin{figure}[]
    \centering
    \includegraphics[width=.88\textwidth,keepaspectratio]{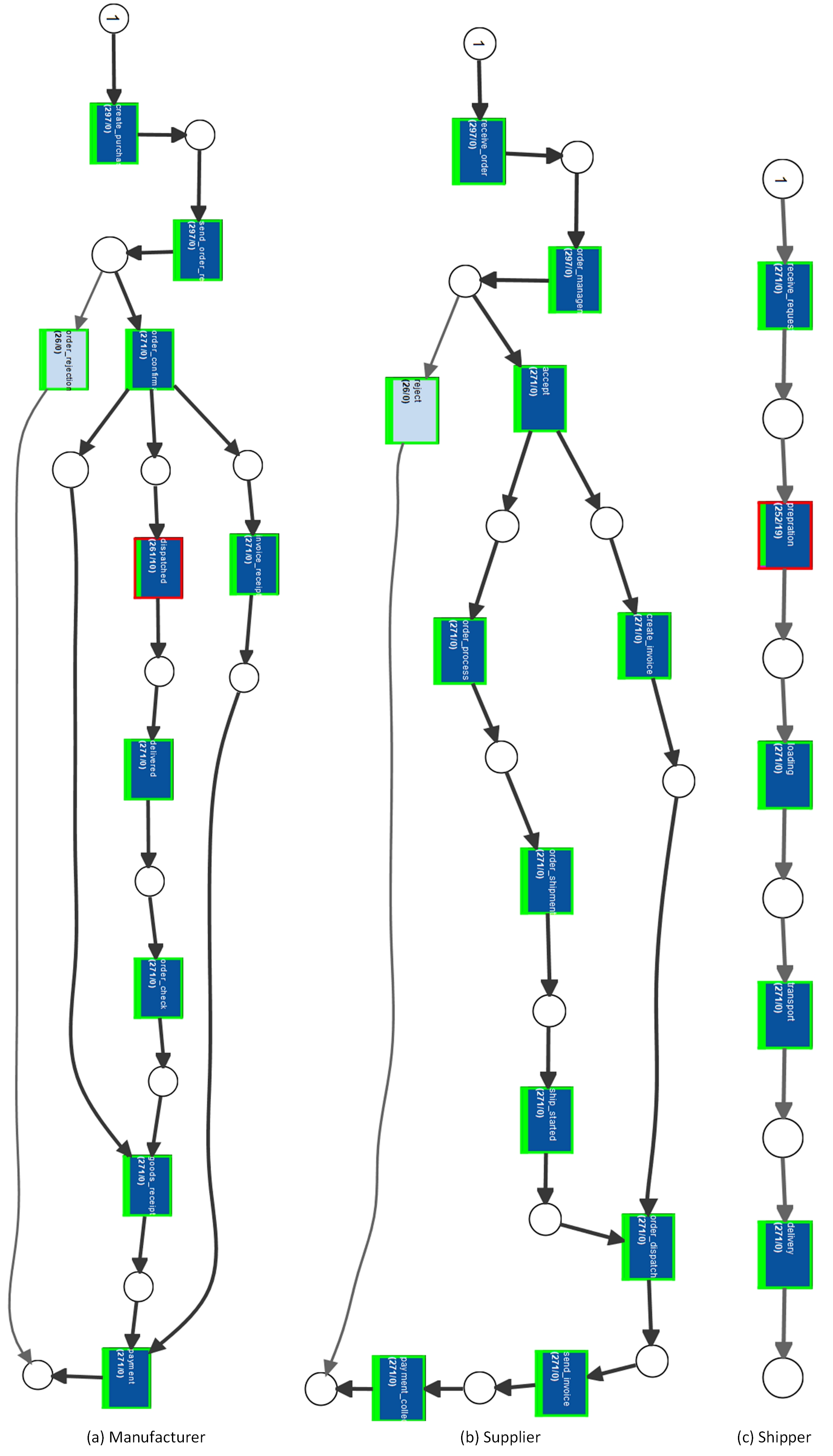}
    \caption{The local alignment results produced by ProM 6.2 for the manufacturer, the supplier, and the shipper. A red border for a transition indicates a model move.}
    \label{fig:local_aligns}
\end{figure}

\begin{sidewaysfigure}[htbp]
    \centering
    \includegraphics[width=1.027\textwidth]{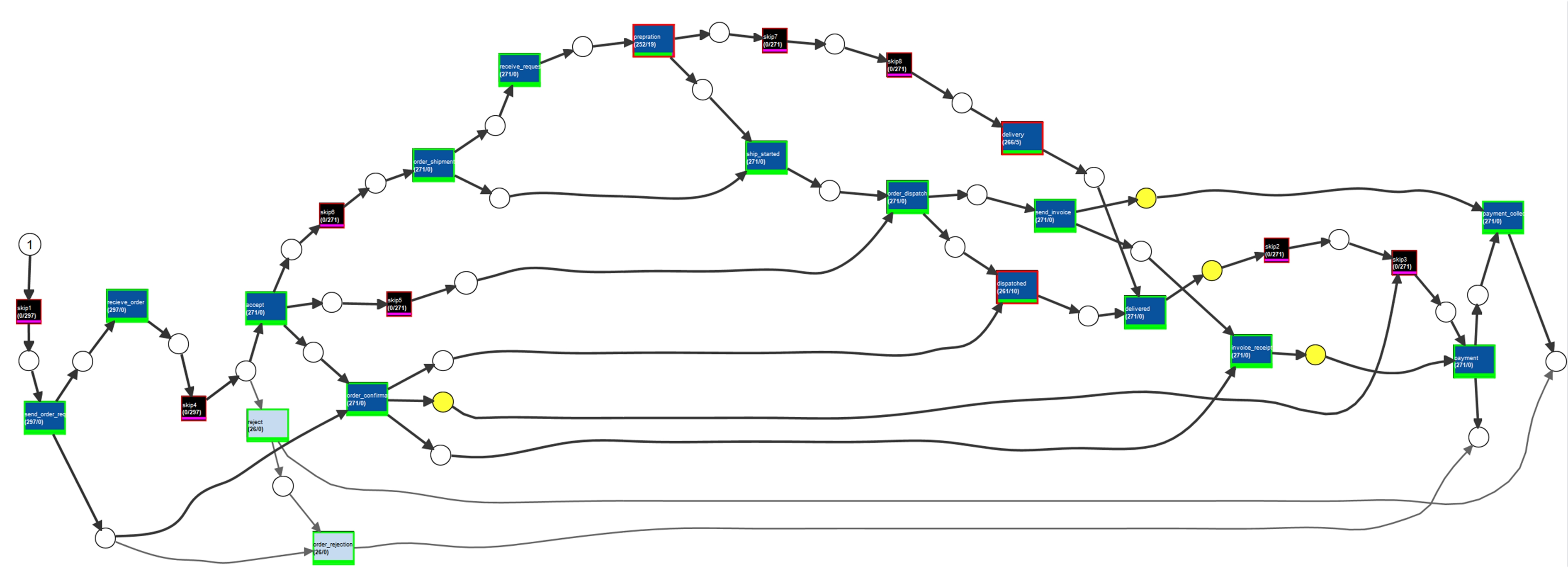}
    \caption{The alignment result produced by ProM 6.2 for the collaborative log and model. A red border for a transition indicates a model move. A yellow place indicates a log move.}
    \label{fig:overall_align}
\end{sidewaysfigure}

\subsection{Local and Federated Conformance Checking}
As shown in Figure~\ref{fig:federated_cc}, the first step is to compute the local alignments and communication costs by each organization. We use no-cost by default for input and output labels associated with the communicational transitions. Thus, potential communication costs associated with the communication points are equal to 1, which refers to a log move, for all the traces where the communication points get involved.

Figure~\ref{fig:local_aligns} shows the local alignment results produced by ProM 6.2 for the manufacturer, the supplier, and the shipper. 
The manufacturer can see a model move on the dispatched activity in its local view due to the cases for which the dispatched activity was not executed, which refers to the sender move scenario.
The supplier sees no misalignment in its local view, even though it involves in two miscommunication scenarios, i.e., the sender and receiver moves.
The shipper can see a model move on the preparation activity in its local view due to the cases for which the preparation activity was not executed, which refers to the receiver move scenario.

As can be seen in the local alignment results, a sender (receiver) move can be reflected in the local alignments of the receiver (sender) organization. However, such local costs cannot be realized as miscommunication without the collaborative log and model.
Moreover, the asynchronous type of miscommunication cannot be reflected in local alignments. This type of miscommunication can only be realized by aligning the collaborative log with the collaborative model.

To perform alignments on the collaborative model and log, we first need to obtain them as described in Definitions~\ref{def:collab_model} and \ref{def:collab_EL}. The collaborative event log and model are available in the GitHub repository\footnote{\scriptsize https://github.com/m4jidRafiei/Federated\_Conformance\_Checking/tree/main/Collaborative}.
Figure~\ref{fig:overall_align} shows the alignment result produced by ProM 6.2 for the collaborative log and model. One can see the sender and receiver moves on the dispatched and preparation activities, respectively. 
Moreover, the asynchronous communication scenario is reflected by 5 model moves and 5 log moves on the delivery and delivered activities, respectively.

Note that in this section, we focused on the deliverable insights provided by the proposed framework that can be obtained by the existing tools rather than calculating the federated alignment costs for each case that has already been shown via the running example.

\section{Conclusion}\label{sec:conclusion}
In conclusion, this paper presented a novel contribution to the field of process mining by addressing the critical aspect of conformance checking in the context of federated (inter-organizational) settings. Our proposed privacy-aware federated conformance checking approach offered a comprehensive framework for evaluating correctness and uncovering non-compliance instances across organizational boundaries.

The value of our approach lies in its ability to provide insights into process compliance issues while respecting the privacy concerns of individual organizations. We introduced public and private models and logs and showed how to compute federated alignment costs by only sharing public models and logs to protect sensitive information, allowing collaborating entities to share and analyze cross-organizational process data without compromising confidentiality.

The empirical evaluation conducted in this paper underscored the efficacy of our approach. Through the design and simulation of a supply chain process involving three distinct organizations, we generated synthetic event logs and showed how our proposed approach could be employed to assess and highlight miscommunications. These insights not only shed light on potential inefficiencies and deviations but also guided process improvement initiatives among collaborating organizations, fostering enhanced coordination and performance in the inter-organizational context.

%
% ---- Bibliography ----
%
% BibTeX users should specify bibliography style 'splncs04'.
% References will then be sorted and formatted in the correct style.
%
\bibliographystyle{splncs04}
\bibliography{refs}

\end{document}